\documentclass[apjl]{emulateapj}

\usepackage{graphicx}
\usepackage{wasysym}
\usepackage{amsbsy}
\usepackage{natbib}
\usepackage[colorlinks,urlcolor=cyan,citecolor=blue,linkcolor=blue]{hyperref} 

\slugcomment{Accepted to ApJL on 15 June 2016}

\shortauthors{} 
\shorttitle{}
\begin{document}

\title{Probing TRAPPIST-1-like systems with {\it K2}}

\author{
Brice-Olivier Demory\altaffilmark{1}, Didier Queloz\altaffilmark{1}, Yann Alibert\altaffilmark{2}, Ed Gillen\altaffilmark{1}, Michael Gillon\altaffilmark{3}
}

\altaffiltext{1}{Astrophysics Group, Cavendish Laboratory, J.J. Thomson Avenue, Cambridge CB3 0HE, UK. bod21@cam.ac.uk}
\altaffiltext{2}{Physikalisches Institut \& Center for Space and Habitability, Universit\"at Bern, 3012 Bern, Switzerland.}
\altaffiltext{3}{Institut d'Astrophysique et de Geophysique, Universite de Liege, Allee du 6 Aout 17, Bat. B5C, 4000 Liege, Belgium.}

\begin{abstract}
The search for small planets orbiting late M dwarfs holds the promise of detecting Earth-size planets for which their atmospheres could be characterised within the next decade. The recent discovery of TRAPPIST-1 entertains hope that these systems are common around hosts located at the bottom of the main sequence. In this Letter, we investigate the ability of the repurposed {\it Kepler} mission ({\it K2}) to probe planetary systems similar to TRAPPIST-1. We perform a consistent data analysis of 189 spectroscopically confirmed M5.5 to M9 late M dwarfs from campaigns 1-6 to search for planet candidates and inject transit signals with properties matching TRAPPIST-1b and c. We find no transiting planet candidates across our {\it K2} sample. Our injection tests show that {\it K2} is able to recover both TRAPPIST-1 planets for 10\% of the sample only, mainly because of the inefficient throughput at red wavelengths resulting in Poisson-limited performance for these targets. Increasing injected planetary radii to match GJ\,1214\,b's size yields a recovery rate of 70\%. The strength of {\it K2} is its ability to probe a large number of cool hosts across the different campaigns, out of which the recovery rate of 10\% may turn into bona-fide detections of TRAPPIST-1 like systems within the next two years.
\end{abstract}

\keywords{planetary systems - techniques: photometric}

\section{Introduction}

The recent discovery of TRAPPIST-1 \citep[hereafter T-1][]{Gillon:2016} from a relatively small target sample (N$\sim$50) suggests that small planets are frequent around late M dwarfs (hereafter LMD, M5.5 to M9 spectral type). This discovery indeed confirms that an untapped population of small planets exists around late M hosts, similar to what has been expected from core accretion models \citep{Payne:2007,Alibert:2011}. The interest in small planets orbiting LMD is justified by their enhanced detectability compared to solar analogues and the unprecedented opportunity they offer for atmospheric characterisation with near-to-come facilities such as JWST \citep{Seager:2014}. 

While LMD are abundant, they are faint in the {\it Kepler} bandpass \citep[5\% throughput at 900nm;][]{Koch:2010a}, which limits {\it Kepler}'s ability to observe them with high precision. Another complicating factor is {\it Kepler} data cadence of 30 min used for the vast majority of targets, which is similar to typical transit durations of short-orbital period planets orbiting LMD. The large collecting area and quasi-continuous monitoring of about 80 days per campaign counter-balance, to some extent, the quenching at longer wavelengths and limited time-sampling of the transits.

\citet{Demory:2013a} proposed an effort to search for Earth-size planets orbiting LMD and brown dwarfs with {\it K2}. In this Letter, we reflect on 1.5 years of {\it K2} data and the discovery of T-1 to examine how {\it K2} contributes to the search of planets orbiting LMD. More specifically, we study the sensitivity of {\it K2} to planetary systems similar to T-1 and put constraints on the properties of the population of planets orbiting hosts located at the bottom of the main sequence.

\section{Observations}
\label{obs}

{\it K2} has been observing LMD since Campaign 0. Each corresponding field of view is located close to the ecliptic to mitigate drifts of the telescope boresight due to differential solar radiation pressure \citep{Howell:2014}. {\it K2} datasets are unprecedented for LMD because they provide a unique opportunity to search and characterise variability patterns over long timescales compared to the relatively fast rotation period of LMD \citep[e.g.][]{Reiners:2010,West:2015}.

\begin{figure}
\centering
   \epsscale{1.2}\plotone{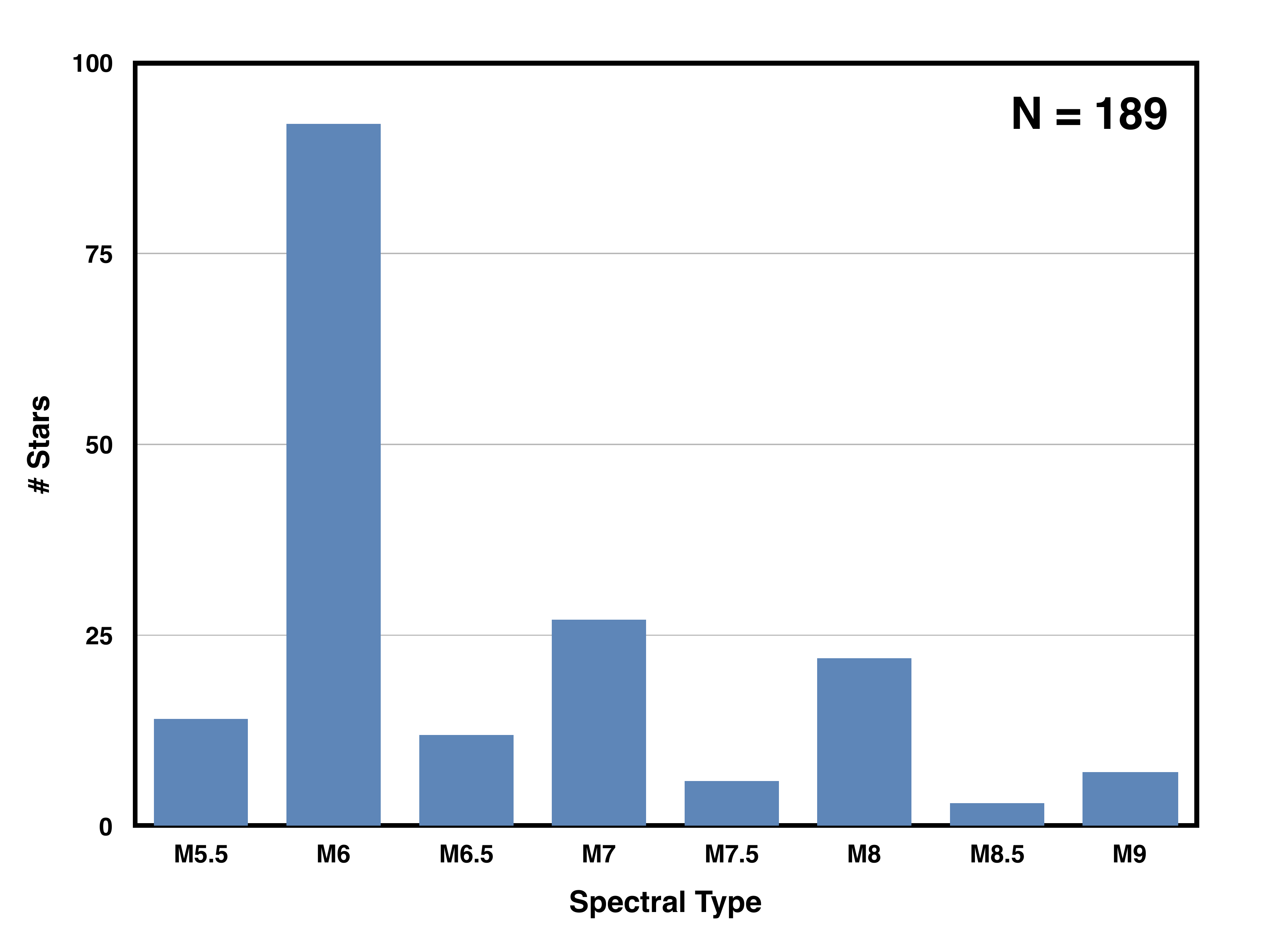}
  \caption{{\bf Spectral type distribution of our sample $\boldsymbol{\it K2}$ LMD.} This histogram shows the spectral distribution of targets that are part of our sample during Campaigns 1 to 6.}  
\label{fig:spt}
\end{figure}

We base this Letter on a total of 189 M5.5-M9 stars that have been observed in K2's Campaigns 1-6, obtained between March 2014 and September 2015. The magnitudes in the {\it Kepler} bandpass range from 14.5 to 23.9. We select only those targets that are confirmed spectroscopically, which allows us to constrain better the host properties for the purpose of transit searches and alleviate contamination from false positives. This selection is done by cross-matching the available {\it K2} targets with spectroscopically confirmed sources from \citet{Cruz:2002,Cruz:2003,Cruz:2007,Lodieu:2007,Reid:2007,Reid:2008,Slesnick:2006,Slesnick:2008,West:2008}. We use this spectral classification to estimate the radius of each target using evolutionary models \citep{Baraffe:2015}. Figure~\ref{fig:spt} shows the distribution of our target sample as a function of spectral type.

\section{Data Analysis}
\label{analysis}

\subsection{Data Reduction}
\label{reduction}

We describe in this section how we perform the data reduction of the {\it K2} data. Our pipeline uses the {\it K2} pixel-level files (\textsc{TPF}) as input data. The \textsc{TPF} files are downloaded from the Mikulski Archive for Space Telescopes (MAST\footnote{\url{http://archive.stsci.edu/k2/}}). Each \textsc{TPF} file is a datacube that includes all frames for a given target. Each frame contains all pixels included in the mask for a given target. Because of the telescope jitter and repointing, each frame encompasses significant margins along the $x$ and $y$ axes to keep the target in the aperture. As the LMD observed within this programme are faint and the masks large enough to include other stars (in addition to the ones we are interested in), our code first identifies the location of the star on the frame based on the target coordinates. We then use a PSF centroiding algorithm on each frame to precisely locate the target on the detector, which will be useful in a second stage for mitigating the photometric systematics. We finally perform simple aperture photometry on individual frames, with star and sky apertures centred on the target position determined in the previous step. We test different aperture sizes and use a portion of the light curve to measure the RMS and level of correlated noise \citep[e.g.][]{Demory:2016} to determine the aperture that minimises both quantities. We eliminate outliers based on datapoints having significant positive median absolute differences \citep{Hoaglin:1983} from the median. The remainder of the data are then ingested into a Markov Chain Monte Carlo (MCMC) framework that includes a polynomial detrending from the centroid position \citep{Demory:2011} previously measured, as well as longer-term trends that are of instrumental or astrophysical origin. The MCMC fit computes uncertainties that include the contribution from correlated noise in the data \citep{Winn:2008b,Gillon:2010a} and errors in coefficients used in the polynomial detrending. The photometric RMS is then computed on the final, detrended light curve. We perform further attempts to improve the light curve extraction by using custom, non-circular-shaped apertures for some targets. A non-circular aperture avoids inclusion of pixels that contains mostly background noise. For this step, we compute for each target a baseline flux value that corresponds to the median of the pixel fluxes where no star is detected, thus containing only background signal. We then define the photometric aperture by selecting only the pixels located at the target location with flux values larger than  the median background flux multiplied by a coefficient that is based on the target's magnitude. We find this approach results in a slightly better photometric RMS for the fainter objects of our sample. We finally compare our photometric reduction with publicly available corrected lightcurves \citep[K2SC, K2SFF, K2VARCAT;][]{Aigrain:2016,Vanderburg:2014,Armstrong:2016a}, all available from MAST. We randomly select 10 targets across our sample and find the photometric RMS to be consistent between these three methods and ours.

\subsection{Planet search}
\label{search}

We then perform a transit search using a box least-squares fitting algorithm \cite[BLS;][]{Kovacs:2002}. We set the BLS to orbital periods ranging from 0.8 days to half of the duration of the Campaign ($\sim$ 40 days). The ratio of the transit duration over the planet orbital period is further set between 0.0007 and 0.06 to include a wider range of orbital periods, eccentricities and impact parameters than the T-1bc planets. We note that the transit duration of planets orbiting LMD can be as short as 20 min for 1-day orbital periods. Because of the {\it Kepler} 30-min cadence photometry, these transits appear significantly smeared out. Long cadence data could thus potentially hamper the detectability of close-in planets with short transit durations. Our pipeline returns for each star the raw/corrected photometry, the stellar motion over the entire campaign duration, and the transit search results. 

Among the 189 LMD that are part of our sample, our BLS analysis yields no transiting candidate detection at a 4-$\sigma$ detection level or above. We explore the possible reasons for this null result below.

\section{Transit injection tests}
\label{injection}

In the following we explore whether the null result regarding our transit search is due to an insufficient sensitivity of K2 to planets transiting LMD. More specifically, in the light of the recent discovery of T-1, we investigate whether our dataset is able to put constraints on the frequency of tight planetary systems orbiting LMD.

\subsection{Recovery of TRAPPIST-1-like systems}

We first model a lightcurve of T-1b (P=1.51d, R$_P$= 1.11 R$_{\oplus}$) and c (P=2.42d, R$_P$= 1.05 R$_{\oplus}$) using the system parameters published in \citet{Gillon:2016}. While the transits are modelled using a 30s cadence, we integrate them to the {\it K2} 30-min long-cadence at which the observations are obtained. The T-1bc planets are then injected in the {\it K2} raw photometry, just after the flux extraction from the {\textsc TPF} files. The transit depths are multiplied by a factor that is the squared ratio of stellar radius between T-1 and the target. We then perform the different steps detailed in Sect.~\ref{reduction} with the exact same sequence. The goal of this part of the analysis is to assess how frequently we detect T-1b and c planets, together or separately. We run the BLS analysis a first time, identify the signal with the maximum signal-to-noise ratio in the periodogram, remove the corresponding data from the corrected phased-lightcurves and run a second BLS analysis. We consider each planet to be successfully recovered if the deduced orbital period matches the input model transit data at 2\% or less. 

Across the 189 systems that are part of our sample, we find that we are able to recover T-1b and T-1c for 20 (11\%) and 12 (6\%) targets respectively. For most cases, the Poisson noise is too large to enable a clear detection of the planets injected in the photometry, leading to false alarms. We find as well that, as expected, the depth recovered by the BLS is impacted by the data sampling. For both T-1b and c, a single photometric point only is located in-transit, and since the orbital period of the planet is not an exact multiple of 30-min, the in-transit point shifts in phase, hence its apparent depth. This reduces the peak power in the BLS spectrum and artificially reduces the strength of the planetary signal.

\subsection{Sensitivity to ``inflated'' TRAPPIST-1-like systems} 
We repeat the same analysis as above but injecting this time planets with radii that are 1.5 and 2$\times$ larger than the T-1bc planets, while retaining all other orbital properties. We find that for a 1.5 scale factor, the recovery rate is 31\% for T-1b and 19\% for T-1c. Injecting transits of planets that are twice as large yields recovery rates of 56 and 40\% for T-1b and T-1c respectively. 

We finally conduct injection tests of planets similar to GJ\,1214\,b in size, using a published radius of 2.68 R$_{\oplus}$ \citep{Charbonneau:2009} for both T-1b and T-1c. We find in that case recovery rates of 71\% and 67\% for T-1b and T-1c respectively. 

\begin{figure}
\centering
   \epsscale{1.2}\plotone{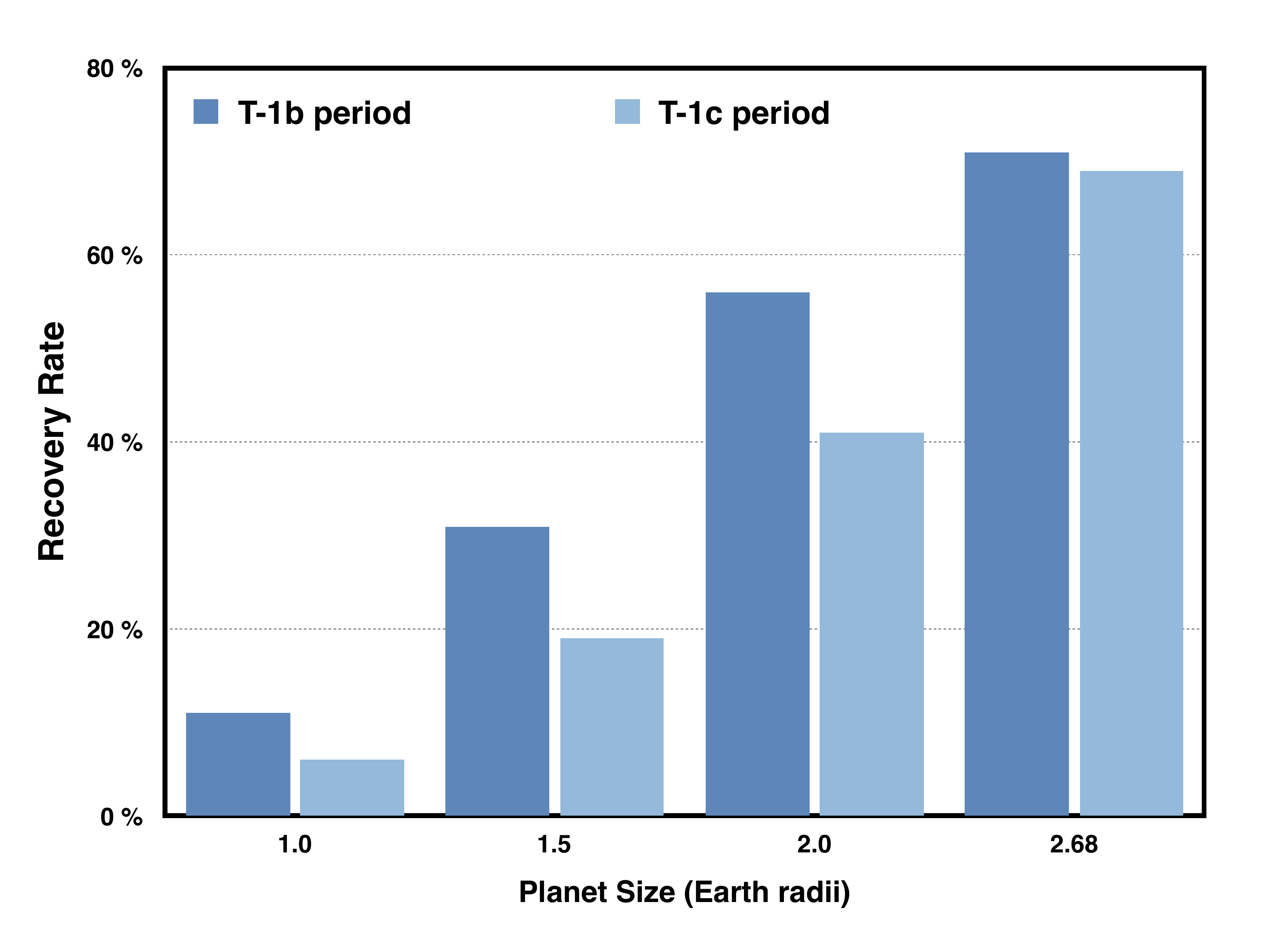}
  \caption{{\bf Recovery rates for TRAPPIST-1 planets.} This histogram shows the percentage of recovery for planets that have been injected in the LMD that are part of our sample. The dark and light bars are the recovery rates for planets with the orbital periods of T-1b and T-1c respectively, assuming radii spanning 1 to 2.68 R$_{\oplus}$.}  
\label{fig:inj}
\end{figure}

\begin{figure*}
\centering
   \epsscale{1.0}\plottwo{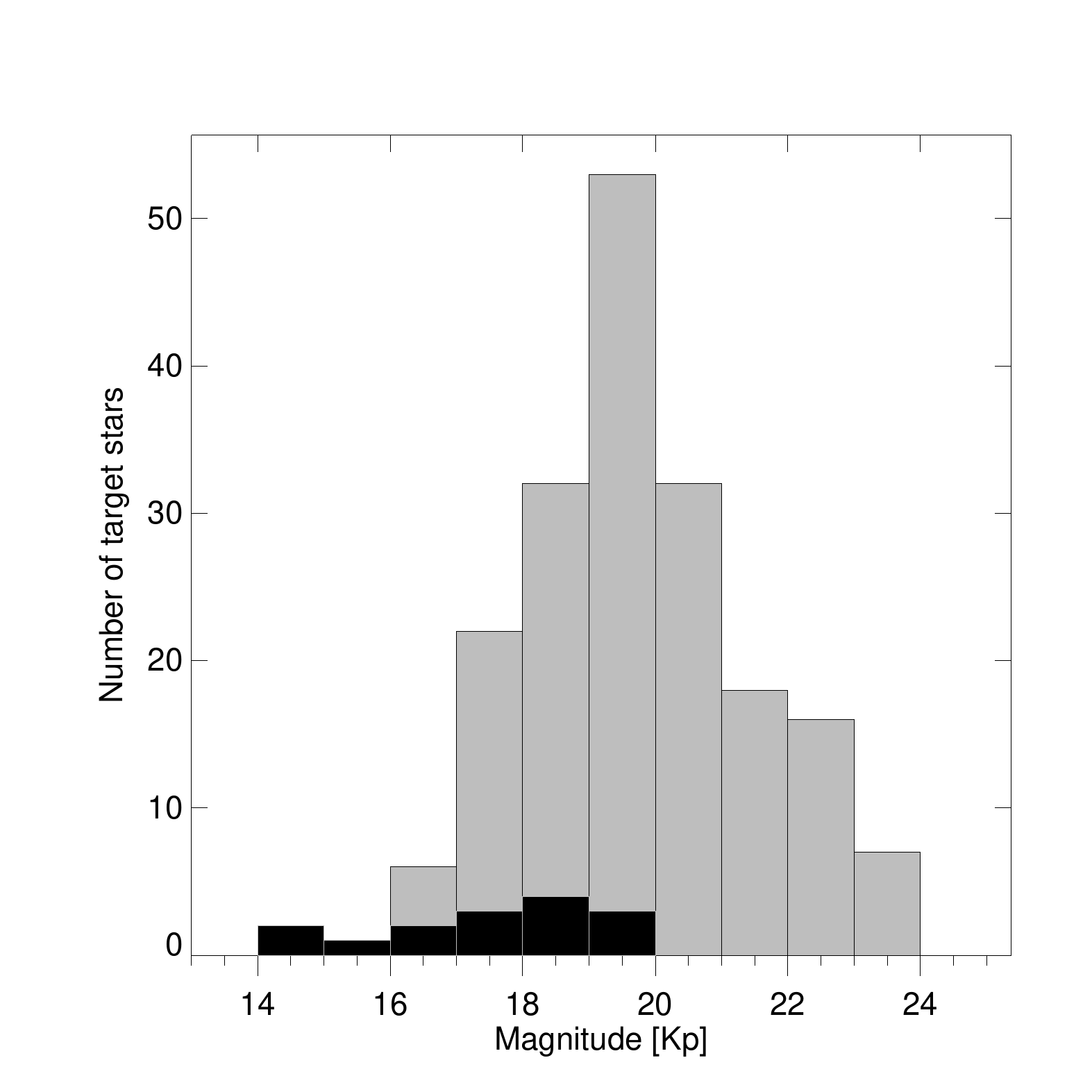}{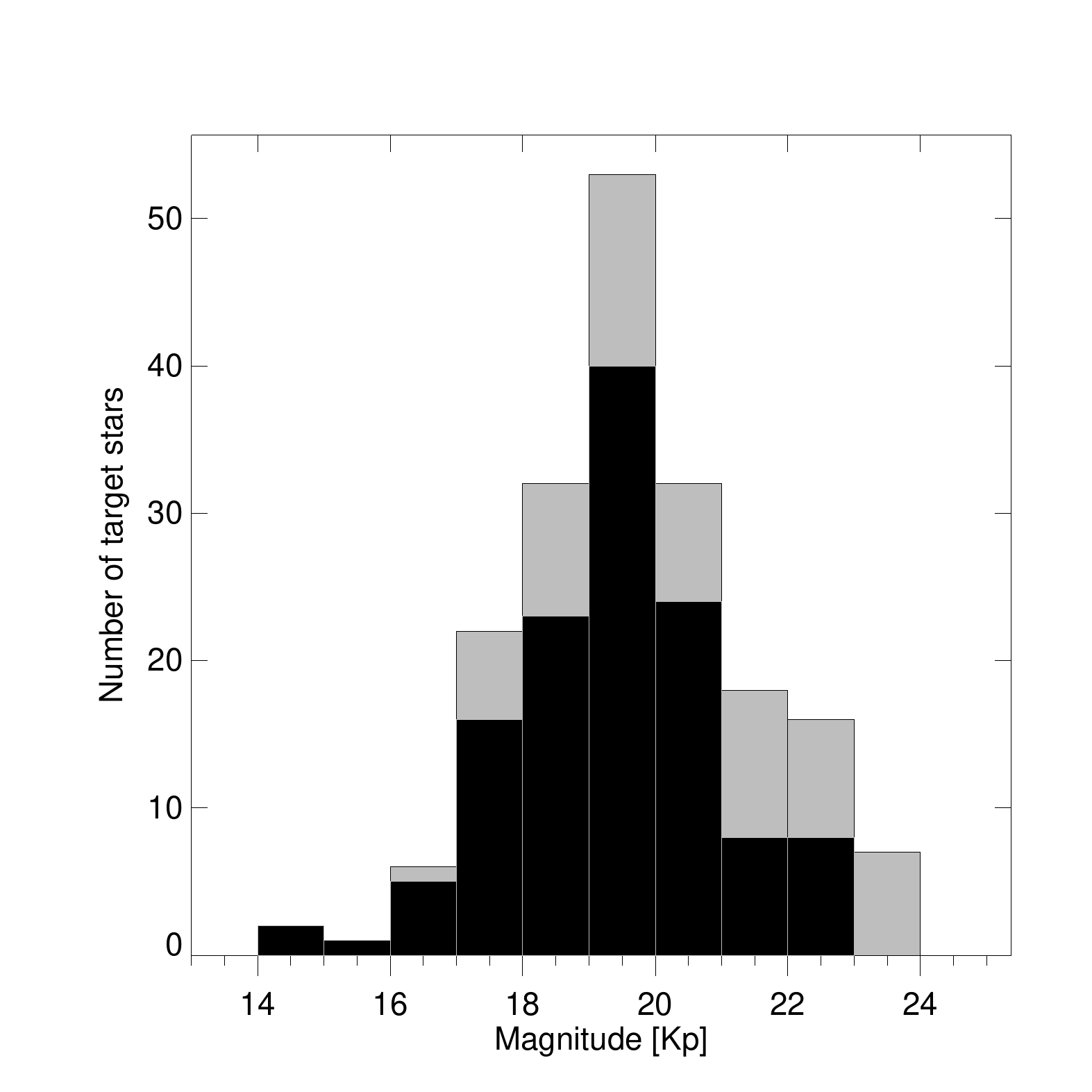}
  \caption{{\bf Detectability of TRAPPIST-1b-like planets.} Distribution of targets included in our sample (grey). Left: distribution of targets for which the injected T-1b signal is correctly recovered (black). Right: distribution of targets for which the injected planet with a period equal to T-1b but with R$_P$=2.68R$_{\oplus}$ is correctly recovered (black).}  
\label{fig:distrib}
\end{figure*}

All recovery rates are shown in Fig.~\ref{fig:inj} for T-1b and T-1c. We show in Fig.~\ref{fig:distrib} the Kp magnitude distribution of target stars (grey) and superimpose the subsample for which the signal is recovered (black) for a planet with a size equal to T-1b (left) and GJ\,1214\,b (right). 

\section{Planet formation models} 
\label{models}

We have computed planet formation models in the framework of the core-accretion scenario, focussing on LMD (0.1 M$_\odot$). Our model is based on the models of \citet{Alibert:2013} and \citet{Fortier:2013}. They both take into account the structure of the protoplanetary disk and its evolution, the migration of the planet, as well as the computation of the planetary growth and composition. 

The mass of the star is included in the models via different aspects \citep{Alibert:2011}, the most important being the distribution of the disk masses, which is different from the one used for solar-type stars, and follows the relation

\begin{equation}
M_{\rm disk} \propto M_{\rm star}^{1.2}
\end{equation}

\begin{figure}
\centering
   \epsscale{1.0}\plotone{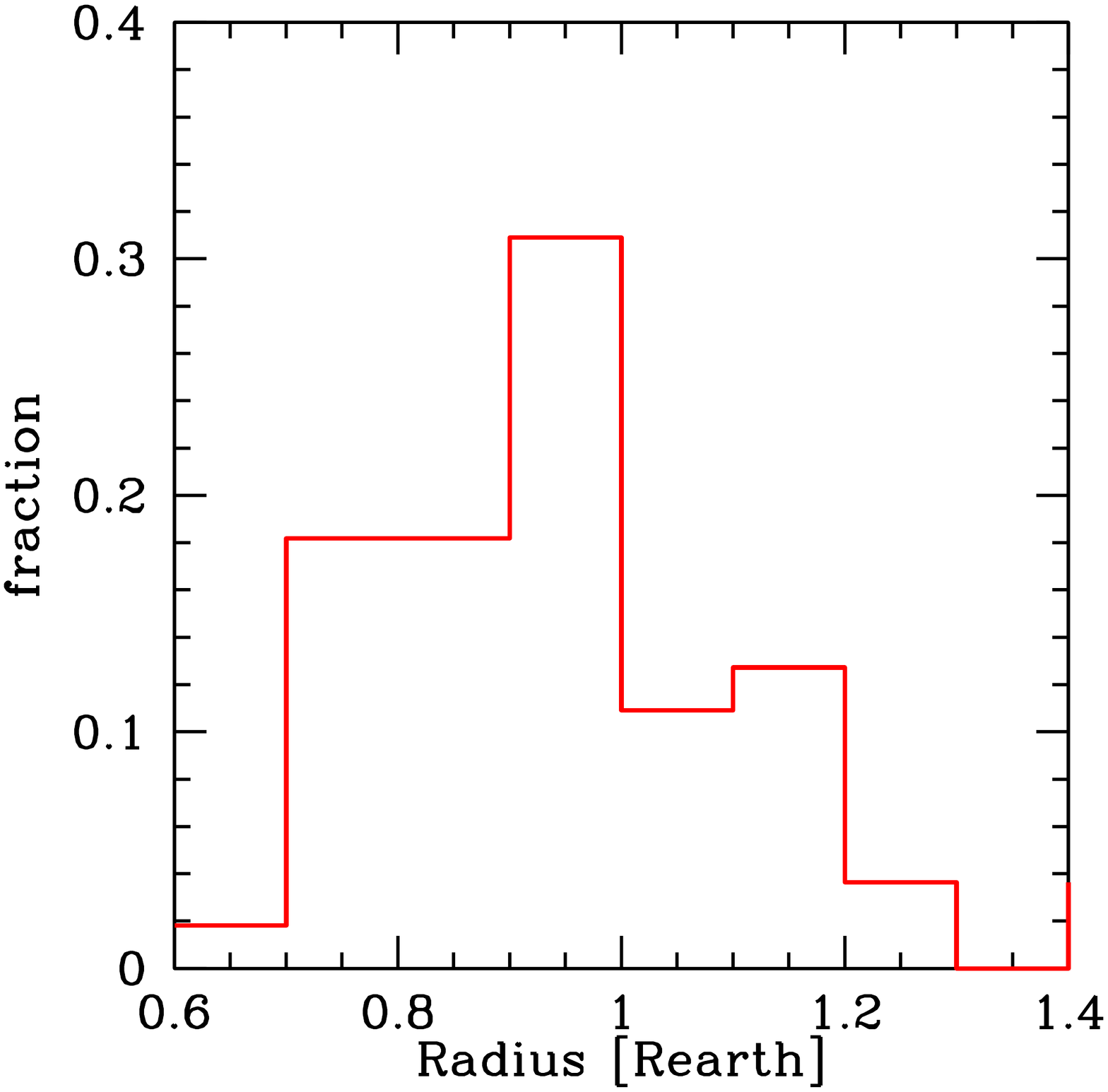}
  \caption{{\bf Planetary radii for planets orbiting a T-1-like star from the formation models of \citet{Alibert:2013} and \citet{Fortier:2013}.} Distribution of planet radii, for planets located inside 0.1 AU of the central star. The radius is computed using the method presented in \citet{Alibert:2014}.}  
\label{fig:models}
\end{figure}

Figure~\ref{fig:models} shows the radius distribution resulting from our models, for planets inside 0.1 AU. As can be seen on Fig.~\ref{fig:models}, the distribution of radii extends from $\sim 0.5$ to $\sim 1.4$ R$_{\oplus}$, with the majority of planets having a radius of the order of 1 R$_{\oplus}$ or smaller. These values are similar to the ones of the recently discovered T-1bcd planets \citep{Gillon:2016}, and we find no planet to be formed with a substantially larger radius. As shown in Fig.~\ref{fig:models}, the recovery rate for these synthetic planets is expected to be small (of the order of 20 \% maximum).

Some of these planets harbour a non negligible fraction of water, because they start their formation process beyond the iceline (the iceline is located between $\sim$0.2 and $\sim$1 AU, depending on the mass of the protoplanetary disk). Interestingly, precise enough determination of the planetary bulk density can allow the determination of the fraction of volatiles, for the shortest period planets. Indeed, if evaporation is efficient enough, short-period low-mass planets cannot retain a gas atmosphere, and the degeneracy in the
determination of the planetary composition is reduced \citep{Alibert:2016}.

\section{Discussion}
\label{discussion}

\subsection{K2 sensitivity to TRAPPIST-1-like systems}

The first conclusion of this Letter is that across the 189 LMD that are part of our sample, we would have been able to detect the T-1 b and c planets for only $\sim$10\% of the targets. Our pipeline efficiently takes correlated noise and stellar variability into account. The dominant source of noise is mostly white and points to the limited performance of {\it K2} to obtain precision photometry for cool and faint LMD.

Assuming that each star was hosting such a system (occurrence rate of 1), and assuming a 100\% recovery rate and a geometric transit probability of $\sim$5\% for these systems, we should have detected about 5 systems per 100 targets. However, the fact that our recovery efficiency on this sample for T-1-like systems is only 10\% means that we would have recovered 0.95 systems across our sample, which is consistent with our non-detection. We estimate that a total of $\sim$1000 well characterised LMD will be observed by the end of 2018 by {\it K2}, if the mission is extended at the next NASA senior review. If the trend in occurrence rate of planets orbiting {\it Kepler} early M stars \citep{Dressing:2013} extends into the LMD regime, we should expect {\it K2} to be able to detect up to $\sim$5 of these Earth-size systems within the next 2 years. However, if the occurrence rate of TRAPPIST-1-like systems is lower than 20\%, no detections from {\it K2} are expected.

Figure~\ref{fig:distrib} (left) shows that {\it K2} sensitivity to T-1b-like planets orbiting LMD extends to Kp$\sim$20, and the recovery rate decreases in a Poisson-limited regime down to the tail of the sample magnitude distribution (i.e. Poisson-limited performance yields no detections for Kp$>$20.

\subsection{A possible lack of close-in super-Earth-size exoplanets orbiting late M dwarfs}

The second conclusion of this Letter is that despite our ability to detect companions larger than the T-1bc planets the orbiting LMD, we find no objects that are 1.5 times the size of the T-1b or larger.
When injecting T-1bc planets with 1.5 times the size of the Earth (super-Earth regime), we find that these planets are fully recovered for $\sim$31\% of the target sample for orbital periods similar to T-1b. We should thus have found 31\% recovered $\times$ 189 hosts $\times$ 5.4\% transit probability = 3.2 planets, assuming an occurrence rate of 1. Similarly, when using a size factor of 2.0 rather than 1.5, we would have found 5.7 such ``inflated'' T-1b-like planets within our sample.

The cause for this null result regarding the planet search in the present sample could be multi-fold. First, larger exoplanets may orbit farther from their host, thus reducing their probability of transit significantly, which would render our observations consistent with non-detections. Second, the distribution of the frequency of planets around LMD may peak towards smaller radii than it does for more massive hosts, as suggested from planet formation models (Sect.~\ref{models}). Assuming that the ratio of the disk mass to the host star mass is constant from solar to the brown-dwarf regime, as shown from FIR observations of young systems \citep[e.g.][]{Mohanty:2013}, LMD would form mostly small planets. Third, the occurrence rate for these planets may be significantly smaller than 1.

It is worth noting as well that for a given dataset we have 1.6 times the number of transits of T-1b compared to T-1c and both have similar depths. Thus, we would expect, all other things being equal, to have a ratio in recovery rates of 1.3. We find that this factor does not fully explain the discrepancy in recovery rate for our injection tests of Earth-size planets but accounts for the difference in recovery ratio between T-1b and T-1c for the 1.5 and 2.0 Earth radii particularly well (see Sect.~\ref{injection}). This suggests that the lightcurves are dominated by Poisson statistical noise, except for the lower amplitude signals, where residual correlated noise of astrophysical or instrumental origin likely complicate the retrieval of the individual transits. Another useful check is to compute the ratio of recovered signals vs. the planetary sizes. For 1.5 and 2.0 size factors, the corresponding transit depth is 2.25 and 4 times larger. From our T-1b recovery rates of 11, 31 and 56\%, the corresponding factors are 2.8 and 5.1, which are both in good agreement. For T-1c, the recovery rates are  6, 19 and 40\%, yielding factors of 2.7 and 5.8, suggesting that the shorter period of T-1b makes its recovery easier especially for small planetary sizes.

\subsection{A lack of close-in mini-Neptune exoplanets orbiting late M dwarfs}

The third conclusion of this paper is that while {\it K2} has excellent sensitivity to mini-Neptune exoplanets, similar in size to GJ\,1214\,b, none are found in our sample. Our injection tests confirm, however, that these objects would have been found in our sample and we would have expected to find 71\% recovered $\times$ 189 hosts $\times$ 5.4\% transit probability = 7.2 planets, assuming an occurrence rate of 1 planet per star. Our findings suggest that the occurrence rate of mini-Neptunes orbiting LMD is likely an order of magnitude smaller at least, making them rare around this population of targets, similar to early-/mid- and late-M stars as shown by observations \citep{Dressing:2013,Berta:2013} and models (Sect.~\ref{models}) respectively. Contrary to T-1b-like planets, Fig.~\ref{fig:distrib} (right) shows that the detectability of GJ\,1214\,b-like planets orbiting LMD extend to Kp$\sim$23 and that the limitation is not from Poisson noise anymore. Rather, the recovery rate is almost constant between Kp$\sim$16 and 21 and drops afterwards. We find that no GJ\,1214\,b-like planet is recovered beyond Kp$\sim$23, which is consistent with our sensitivity for T-1b-like planets discussed at the beginning of this Section. The origin of this plateau is thus due to a non-optimal recovery rate at the mid-magnitude range of our sample before Poisson-limited noise dominates again, possibly due to a larger activity level of a sub-sample of our targets.

\begin{figure}
\centering
   \epsscale{1.2}\plotone{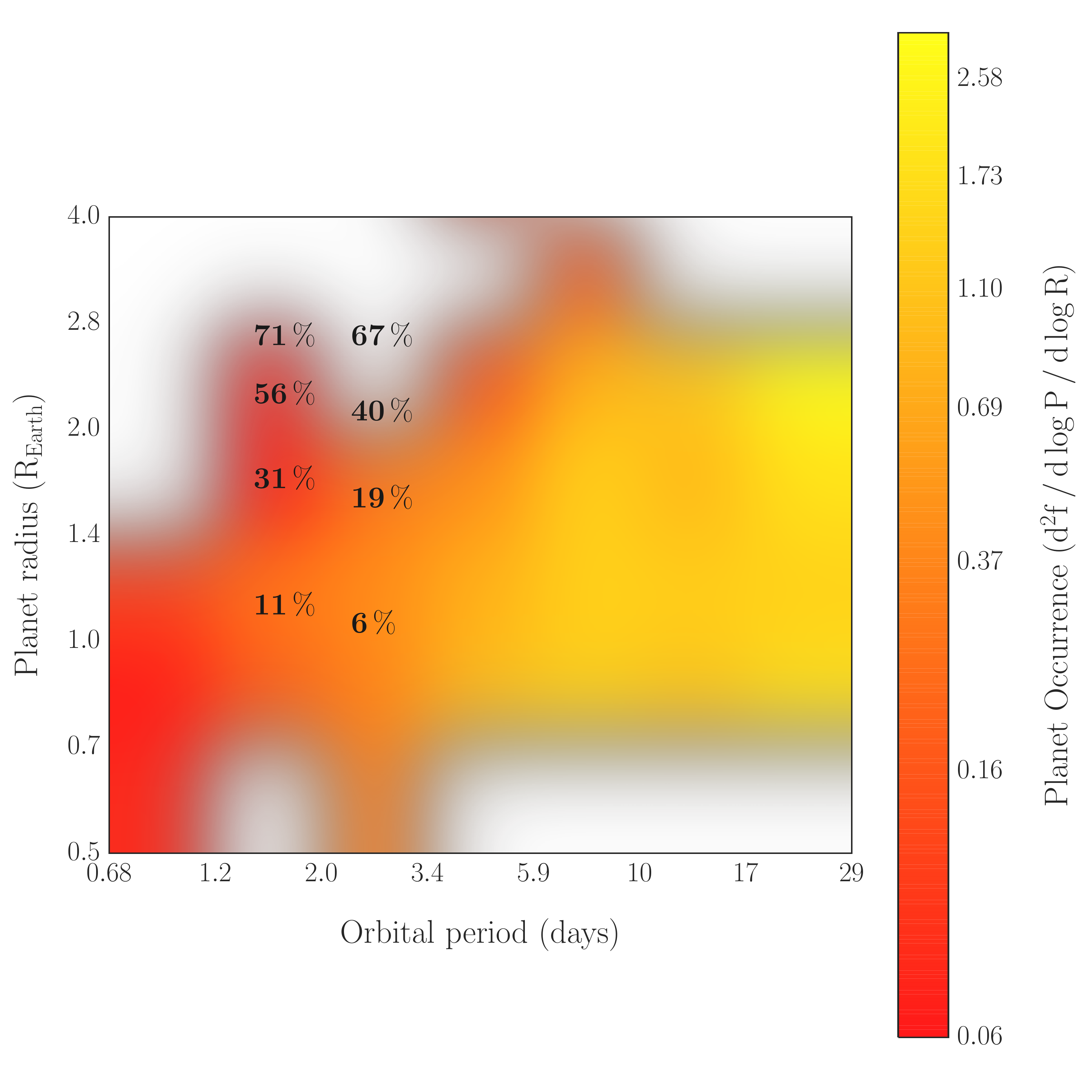}
  \caption{{\bf Planet occurrence and $\boldsymbol{\it K2}$ recovery rate.} The planet occurrence rate from \citet{Dressing:2013} is shown in coloured scale (low to high, red to yellow). White areas represent the absence of detections in the corresponding period-radius bins. The recovery rate (in percent) of planets injected in our sample LMDs at the orbital periods of T-1b and c are shown superimposed for different planet radii. }  
\label{fig:diag}
\end{figure}

We show a summary of our findings in Fig.~\ref{fig:diag}. This diagram shows the occurrence rate of {\it Kepler} planets published for early- and mid-M stars \citep{Dressing:2013} and the superimposed numbers indicate our recovery rate (in percent) at the orbital periods of T-1b and c for different planet radii. Assuming that this occurrence rate distribution extends into the LMD regime, Fig.~\ref{fig:diag} shows that {\it K2} is relatively inefficient at detecting planets with radii $R_p \lesssim 1.4 R_{\oplus}$, where the occurrence rate peaks for 1.51 (T-1b) and 2.42 (T-1c) days. However, formation models produce only few planets at separations smaller than $<$0.05 AU (Fig.~\ref{fig:models}). If models are right, it means that there is a strong transition in the occurrence rate of planets between early and late-type M dwarfs and that the discovery of T-1 out of $\sim$50 targets monitored in the TRAPPIST survey had a very low probability of detection. Alternatively, models could be missing important ingredients that future monitoring of similar systems may help to constrain.

{\it K2} relies on its ability to quasi-continuously monitor several thousands of targets over 80 days for each campaign. By the end of 2018, we may expect {\it K2} to discover one or two systems similar to T-1 and a handful of super-Earth systems if their occurrence rate is similar to the one deduced from the {\it Kepler} primary mission's M stars. We finally note that we have proposed {\it K2} observations of TRAPPIST-1 in Campaign 12. Because this target is bright, it is part of the 10\% subsample for which we will be able to precisely characterise the transits of the known planets and search for other companions in this remarkable system. 

\acknowledgments
We thank the anonymous referee for an helpful review. Part of this work has been carried out within the frame of the National Centre for Competence in Research PlanetS supported by the Swiss National Science Foundation. YA acknowledges the financial support of
the SNSF. M. Gillon is Research Associate at the Belgian F.R.S-FNRS.  This paper includes data collected by the K2 mission. Funding for the K2 mission is provided by the NASA Science Mission directorate.

{\it Facility:} \facility{{\it K2}}


\end{document}